# A Hyperspectral Imaging Dataset and Methodology for Intraoperative Pixel-Wise Classification of Metastatic Colon Cancer in the Liver

Ivica Kopriva, *Senior Member, IEEE*, Dario Sitnik, Laura-Isabelle Dion-Bertrand, Marija Milković Periša, Arijana Pačić, Mirko Hadžija, and Marijana Popović Hadžija

*Abstract*—Hyperspectral imaging (HSI) holds significant potential for transforming the field of computational pathology. However, there is currently a shortage of pixel-wise annotated HSI data necessary for training deep learning (DL) models. Additionally, the number of HSI-based research studies remains limited, and in many cases, the advantages of HSI over traditional RGB imaging have not been conclusively demonstrated, particularly for specimens collected intraoperatively. To address these challenges we present a database consisted of 27 HSIs of hematoxylin-eosin stained frozen sections, collected from 14 patients with colon adenocarcinoma metastasized to the liver. It is aimed to validate pixel-wise classification for intraoperative tumor resection. The HSIs were acquired in the spectral range of 450 to 800 nm, with a resolution of 1 nm, resulting in images of 1384×1035 pixels. Pixel-wise annotations were performed by three pathologists. To overcome challenges such as experimental variability and the lack of annotated data, we combined label-propagation-based semi-supervised learning (SSL) with spectral-spatial features extracted by: the multiscale principle of relevant information (MPRI) method and tensor singular spectrum analysis method. Using only 1% of labeled pixels per class the SSL-MPRI method achieved a micro balanced accuracy (BACC) of 0.9313 and a micro F1-score of 0.9235 on the HSI dataset. The performance on corresponding RGB images was lower, with a micro BACC of 0.8809 and a micro F1-score of 0.8688. These improvements are statistically significant. The SSL-MPRI approach outperformed six DL architectures trained with 63% of labeled pixels. Data and code are available at: https://github.com/ikopriva/ColonCancerHSI.

*Index Terms*—Hyperspectral imaging, computational pathology, adenocarcinoma of the colon in the liver, frozen sections, multiscale principle of relevant information, tensor singular spectrum analysis, semi-supervised learning, deep learning.

## I. INTRODUCTION

SEGMENTATION and classification of medical images are crucial for disease diagnosis, therapy planning, follow-up tracking of therapy efficiency, and intraoperative tumor resection. It is well known that life expectancy improves with the extensive resection of both primary tumors and their metastasis, such as colon adenocarcinoma in the liver [1]. This also applies to other tumors, including gliomas [2] and oral squamous cell carcinomas [3]. In particular, the margin of non-tumor tissue surrounding the tumor, known as the resection margin [38], is a powerful predictor of the 5-year survival rate [39] [40] [11]. However, achieving total resection is challenging due to tumor infiltration into surrounding tissue, making tumor borders difficult to identify. Surgeons rely on intraoperative information provided by pathologists, who perform near-real-time analysis of rapidly frozen and stained histopathological sections. Traditionally, as noted in [4], computational pathology has relied on RGB images of these sections, which limits information to the visual range and excludes the data across the continuous spectral range [4]. To address this limitation, hyperspectral imaging (HSI) technology is increasingly being used in various medical imaging applications, including computational pathology. For comprehensive reviews, we refer interested readers to [4] for an overview of computational pathology from 2013 to 2019, to [5]

This paragraph of the first footnote will contain the date on which you submitted your paper for review. This work was supported by the Croatian Science Foundation Grants IP-2022-10-6403 *(Corresponding author: Ivica Kopriva.)*

Ivica Kopriva is with the Division of Computing and Data Sciences, Ruđer Bošković Institute (RBI), Zagreb Croatia (e-mail: ikopriva@irb.hr).
Dario Sitnik is with the Technical University Munich, Munich, Germany (e-mail: dario.sitnik@gmail.com).
Laura-Isabelle Dion-Bertrand is with the Department of Physics, University of Montreal, Montreal, Quebec H2C OB3, Canada (e-mail: laura-isabelle.dion-bertrand@umontreal.ca). This work was done while she was with the Photon etc., Montreal, Canada.
Marija Milković Periša is with the Clinical Hospital Center Zagreb, Zagreb, Croatia and with the Institute of Pathology, School of Medicine, University of Zagreb, Zagreb, Croatia (e-mail: mmperisa@gmail.com).
Arijana Pačić is with the Department of Pathology and Cytology, Clinical Hospital Dubrava, Zagreb, Croatia (e-mail: arijanapacic@yahoo.com).
Mirko Hadžija was with the Division of Molecular Medicine, RBI, Zagreb, Croatia (e-mail: Mirko.Hadzija@irb.hr).
Marijana Popović Hadžija is with the Division of Molecular Medicine, RBI, Zagreb, Croatia (e-mail: Marijana.Popovic.Hadzija@irb.hr ).

for surgical applications over the same period, and to [6] for medical HSI research from 1998 to early 2013.

Although HSI has demonstrated its potential to detect diseases-related pathological changes in unstained histopathological specimens [7], [8], it is predominantly applied to stained histopathological sections [9]-[24]. However, several unresolved issues still prevent the routine use of HSI in computational pathology. While HSI is generally assumed to offer performance improvements over RGB imaging, the number of studies supported this claim is limited, as highlighted in [4], and further research is needed to confirm the advantages of HSI technology over traditional RGB images. Studies such as [20][25] are addressing this gap. Since HSI captures both spatial (morphological) and spectral information from specimens, classifiers that leverage both spectral and spatial features have been shown to outperform those relying solely on spectral information [9] [10] [11] [18] [20] [22] [23] [26] [27] [29] [30]. Most of the architectures cited above are deep learning networks, which require large amounts of annotated HSI data for training to avoid overfitting. However, as noted in [4], there is a significant lack of annotated HSI databases. Two primary factors contribute to this shortage. First, due to time constraints, data collection during intraoperative procedures is demanding, even for RGB microscopic images [1]. Second, pixel-wise annotation requires pathologists, whose availability is limited. Additionally, the appearance of HSI data differs significantly from RGB images, making the annotation process even more challenging [9]. These factors hinder the routine usage of HSI in computational pathology. To address this, several HSI databases have been created [20], [28] [31] [32].

Motivated by the issues outlined above we present herein the following contributions:

1) HSI data base creation: We introduce a novel database of 27 HSIs of hematoxylin-eosin (H&E) stained frozen sections, collected intraoperatively from 14 patients with colon adenocarcinoma metastasized to the liver. The HSIs were acquired in spectral range from 450 to 800 nm, with a 1 nm resolution. In addition to the HSIs, the dataset includes pixel-wise ground truth maps labeled by three pathologists, distinguishing cancerous and non-cancerous pixels, as well as co-registered pseudo-RGB images. To our knowledge, this is the first HSI dataset of frozen sections for this specific type of cancer, addressing the scarcity intraoperative HSI datasets, particularly for metastatic colorectal cancer.

2) Semi-supervised learning (SSL) and spectral-spatial features: To address challenges such as costly expert labeling, experimental variability, and limited annotated data, we propose label-propagation-based semi-supervised learning. In combination with SSL, we leverage two novel spectral-spatial feature extraction techniques: (*i*) the multiscale principle of relevant information (MPRI) [35] that, thanks to the PRI [36], extracts the multiscale information embedded into HSI data by learning representation on a coarse-to-fine manner; (*ii*) the tensor spectrum singularity analysis (TensorSSA) method [64]. Both the SSL-MPRI and the SSL-TensorSSA methods perform well using only 2% (or as little as 1% for SSL-MPRI) of labeled data per class, delivering high-quality classification results on individual HSIs. These methods, unlike deep networks, can be applied on an HSI-by-HSI basis, making them suitable for scenarios with very few HSIs.

3) Outstanding performance, significantly better than the one achieved by deep networks that require much more labeled pixels for training, is obtained with 1% of labeled samples per class only. For this case, the SSL-MPRI demonstrated the micro performance of 0.9313 in balanced accuracy, 0.9235 in Dice coefficient ($F_1$) score, 0.8578 in IoU (Jaccard index) and 0.9076 in precision (positive predicted value). Corresponding results obtained by the nnUnet, [58], deep network averaged after 10 runs, trained on 17 HSIs from 9 patients and tested on 10 HSIs from 5 patients, were 0.8904, 0.8654, 0.7627 and 0.8272.

## II. RELATED WORK

Hyperspectral imaging technology has been widely used for many years in remote sensing, supporting applications in resource management, agriculture, exploration of minerals, monitoring of environment, and more [37]. Since the late 1990s, HSI has also been applied to various medical imaging modalities, showing significant potential in cancer diagnosis for the cervix, breast, colon, head and neck, prostate, ovary and lymph nodes, among others. For more detailed references on these applications see [6].

### A. Hyperspectral imaging in computational pathology

Reference [4] provides an extensive review of the use of microscopic HSI and multispectral imaging in the analysis of histological specimens from 2013 to 2019. The review covers 84 articles across various medical diagnostic fields, including hematology, breast, central nervous system, gastrointestinal, genitourinary, head and neck, and skin. In this paper, we also briefly review recent developments in microscopic HSI. For instances, a semi-supervised nonnegative matrix factorization approach for pixel-wise classification of HSI in unstained colon cancer sections was proposed in [8], requiring only a fraction of the pixels to be labeled during training. The method was validated on HSIs of unstained colon cancer sections. Additionally, a self-supervised spectral regression method was introduced in [9] for the classification of HSIs from stained specimens with pancreatic adenocarcinoma, with the aim of reducing the need for labeled pixels. This method, applied to 523 HSIs of 36 patients (331 training, 101 validation, 91 testing), achieved an accuracy of 92.31±1.10%. Moreover, a 3D convolutional neural network (CNN), named Hyper-net, was proposed in [10] for the segmentation of melanoma HSIs. A HSI was of size 1024×1024 pixels, with a spatial resolution of 3μm at 20× magnification and a spectral resolution of 7.5nm in the range of 500-1000 nm. Band selection was performed based on mutual information criterion, with 16 bands from 670 nm to 783 nm chosen for segmentation task. Ground truth annotations were provided by pathologists using both an RGB image and the 20th band image. The 3D Hyper-net achieved a Dice coefficient 0.91 on the test images. For segmentation of tongue cancer images, HSIs in the 400-1000 nm range were combined with several variations of the U-net architecture [11]. Using a leave-patients-out cross-validation approach with a clinical dataset of 14 patients, this method yielded a Dice coefficient of 0.924±0.036. In [13], a dimensionality reduction method for HSI was proposed for intra-operative tumor margin definition during brain surgery. Additionally, an in-house developed polarization-based HSI system was presented in [14], combined with artificial neural network (ANN) for diagnosing skin complications caused by diabetes mellitus (DM) at an early stage. HSI was able to detect systemic and local microcirculatory changes associated with DM [15]. Finally, a method was proposed in [16] for flat-field correction, addressing the issue of uneven illumination problem in microscopic hyperspectral imaging. In [17] linear discriminant analysis and support vector machine (SVM) classifiers were applied to HSIs containing 60 spectral bands from H&E stained specimens to distinguish elastic and collagen fibers. The classification accuracy was approximately 4% higher compared to results obtained using RGB images. In [18], HSI was explored for the automatic detection of head and neck cancer nuclei in histologic slides and for identifying cancerous regions based on nucleus detection. HSIs were acquired from 20 patients with squamous cell carcinoma and co-registered with RGB images. Spectra-based SVM and patch-based CNNs were trained using GB patches (RGB-CNN) and HSI patches (HSI-CNN), achieving testing accuracies of 0.74 and 0.82. In [19], classification model that combines spectral and spatial information from HSIs was proposed to distinguish cancer from healthy tissue. The method, which included an automated algorithm based on a minimum spanning forest and optimal band selection, was validated on animal model. The SVM-based pixel-wise classifier achieved an overall accuracy of 91.6% with a sensitivity of 98.2%. In [20], a comparative study was conducted to assess deep networks trained on HSIs and RGB images of histopathological slides. The oral and dental spectral database, containing 215 manually segmented spectral images with 35 classes from 30 human subjects, was used. The pixel-wise accuracy for RGB images was 39.49%, while for HSIs it was 49.48%. In [31], evaluation study was conducted to assess the intraoperative margin using HSIs from fresh human breast specimens. The diagnostic performance of HSI on tissue slices was notable: invasive carcinoma, ductal carcinoma *in situ*, connective tissue, and adipose tissue were correctly classified as tumor or healthy tissue with accuracies of 93%, 84%, 70%, and 99%, respectively. On the resection surface, HSI detected 19 of 20 malignancies that, according to the histopathology information, were located within 2 mm of the resection margin. In [22], an important-aware network (IANet) is proposed for the segmentation of microscopic HSIs. This network is designed for multiscale feature extraction and was demonstrated on HSIs of cholangiocarcinoma. The images were 1280×1024 pixels in size with 60 spectral bands ranging from 550 nm to 1000 nm. On independent test set, IANet outperformed several competing methods, achieving an accuracy of 88.20%, an intersection-over-union (IoU) of 66.91%, and a Dice of 79.82%. No results for RGB images were reported. In [23], HSI of histopathological sections is combined with machine learning for intraoperative tumor margin assessment. The HSI has size of 720×540 pixels, with a spectral resolution of 10nm from 550 nm to 1000 nm. RGB images, manually annotated by two pathologists, were co-registered with the HSIs. Using a fine-tuned ResNet, the achieved accuracy on HSIs was 0.76±01.0, with an $F_1$ score of 0.74±0.14, and a sensitivity of 0.48±0.24. Paper [24] describes a system for acquiring HSIs of unstained specimens from head-and-neck cancer (oral cavity) for surgical margin assessment. Spectral curves were averaged over non-overlapping 5×5 blocks, and four types of features were extracted from these block-images. LDA and SVM classifiers were trained to assign either cancerous or normal labels to each block. Performance on 16 subjects for distinguishing between cancerous and non-cancerous tissues achieved an accuracy of 90±8%, a sensitivity of 89±9%, and a specificity of 91±6%. In [27], a transformer architecture modified with a sparsity module to enhance contextual learning was embedded in the encoder part of a u-shape segmentation network. At each spatial location, the input consisted of convolutional features that were combined with spatial information, which was then entangled with spectral information encoded by the transformers. This created a joint spectral-spatial representation learning framework. Experimental results highlighted the importance of this scheme for spectral contextual learning. On a cholangiocelular carcinoma dataset, the proposed spectral transformer achieved a 12% higher Dice coefficient than U-Net and nearly 7% higher than U-Net++. Paper [28] introduces a database containing HSIs from 68 patients with two types of membranous nephropathy (MN). A tensor patch-based linear discriminative regression method was proposed for classification, exploiting the characteristics of medical HSIs. The combination of a tensor-based classifier and hyperspectral imaging offers a new research direction in kidney pathology with clinical potential for automatic MN diagnosis. In [29], a two-stage deep learning framework for semantic segmentation of HSIs of cholangiocarcinoma was proposed. The first stage used Label-to-Photo translation to train a generative adversarial network to generate realistic images from pixel-level annotations. In the second stage, the generator with frozen parameters was employed to generate hyperspectral images for data augmentation. This approach is particularly valuable because deep models for pixel-wise classification of cholangiocarcinoma require precise annotations, which are time-consuming for pathologists, especially given the rarity of the disease and the scarcity of experienced specialists. On Choledoc dataset, the proposed framework achieved a mean IoU of 76.16%, a mean Dice of 85.80%, and a mean accuracy of 90.96%. Reference [31] describes a hyperspectral microscope and the creation of a hyperspectral database consisting of in-vitro human brain tissue samples. SVM and ANN were applied to distinguish between healthy and tumor brain tissue samples, achieving a sensitivity and specificity of over 92%. Reference [32] addresses a key challenge in surgical HSI - the development of public intraoperative HSI neurosurgical database. The HSI camera captures images within the 500 to 900 nm spectral range, with a spatial resolution of 1024×1024 pixels and a spectral resolution of 3 to 5 nm. Corresponding RGB images were also obtained as anatomical references, which were used to label the HSI data. After aligning and scaling the RGB images, two neurosurgeons performed the annotation using Microsoft Paint, labeling healthy tissue, pathological tissue, fluorescence, and irrelevant data (shadows, noise, out-of-focus areas, specular reflections). The database includes 52 HSIs acquired during 10 microsurgical operations. Reference [41] discusses the use of HSI acquired over micro-FTIR HSI absorbance spectroscopy for characterizing of cancerous, inflammatory and healthy colon tissues. In total 71, HSIs were acquired: 24 form healthy tissue, 27 from inflammatory tissue and 20 from cancerous tissue. These HSIs, organized in voxel format, were used to train both a fully connected deep neural network (DNN) and linear SVM for tissue classification. Using a K-fold cross-validation protocol, the DNN achieved an accuracy and sensitivity of 99%, while the linear SVM achieved a sensitivity of 96.48%.

As highlighted in the presented overview, no public database currently exists that provides HSIs of H&E stained frozen sections of adenocarcinoma of the colon in the liver. This gap underscores the significance of our paper. Additionally, the second contribution of this paper is justified by the need to develop an architecture that: (*i*) efficiently extracts discriminative spectral-spatial features from HSIs, (*ii*) reduces the reliance on large amount of annotated pixels for training, (*iii*) adapts to varying experimental conditions, and (*iv*) clearly demonstrates that the use of HSIs offers performance improvements over RGB images.

## III. METHODS

### A. Hyperspectral image database

The database consists of of 27 HSIs of H&E stained frozen sections, collected intraoperatively from 14 patients diagnosed with adenocarcinoma of the colon in the liver. These specimens were gathered through a clinical study funded by the Croatian science foundation grant IP-2016-06-5235, conducted between March 2017 and February 2020 at the Department of Pathology and Cytology, Clinical Hospital Dubrava, Zagreb, Croatia. This dataset is a part of larger cohort of 19 patients with the same diagnosis, for which 82 RGB images with pixel-wise annotations are publicly available at https://cocahis.irb.hr [1]. The Institutional Review Board of Clinical Hospital Dubrava approved the collection of samples on May 24, 2016. All patients provided written informed consent, and the data were anonymized. The HSIs were recorded across a spectral range of 450 to 800 nm, with a spectral resolution of 1 nm, and a spatial resolution of 0.11 μm$^2$. Each image has a size of 1384×1035 pixels and is stored in HDF5 format. Accompanying each HSI is a binary ground truth map created through majority voting of pixel-wise annotations by two pathologists (A.P and M.M.P) and a medical expert (M.H).

### B. Hyperspectral image acquisition

Regions of interest were marked on H&E stained histopathological specimens. These regions were then imaged using Photon etc's hyperspectral fluorescence microscope IMA, [42] [43], as shown in Fig. 1. The IMA employs volume Bragg gratings (VBG) to capture spectrally resolved images, which are combined into a hyperspectral data cube. The use of VBG allows for global imaging, where signals from all points within the field of view are collected simultaneously, avoiding the need for x-y or line scanning. The resulting data cubes are composed of 351 gray scale images, each with a spectral resolution of 1 nm, covering the 450 nm to 800 nm range, with a focus at 590 nm. Broadband source was used for illumination. The images were captured using a 20× objective lens with the exposure time of 500 ms, producing images of 1392×1040 pixels with a field-of-view of 449 μm×335 μm. This corresponds to a spatial resolution of 0.11 μm$^2$ per pixel, which effectively eliminates the mixing of cancer and non-cancer tissues at the pixel level. To avoid edge effects, we retained rows 4:1387 and columns 6:1040 columns, resulting in a final data cube size of 1384×1035×351. Data were normalized using the lamp spectrum. Fig. 2 shows spectral profiles of one pixel labeled as cancerous and another labeled as non-cancerous.

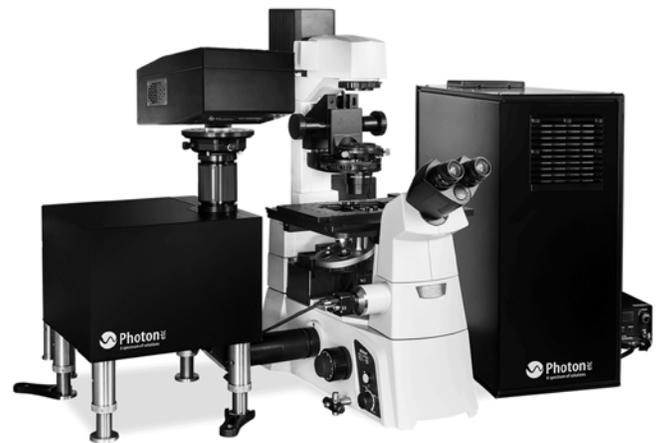

Fig. 1. Hyperspectral fluorescence microscopy system equipped with volume Bragg gratings and an optical microscope.

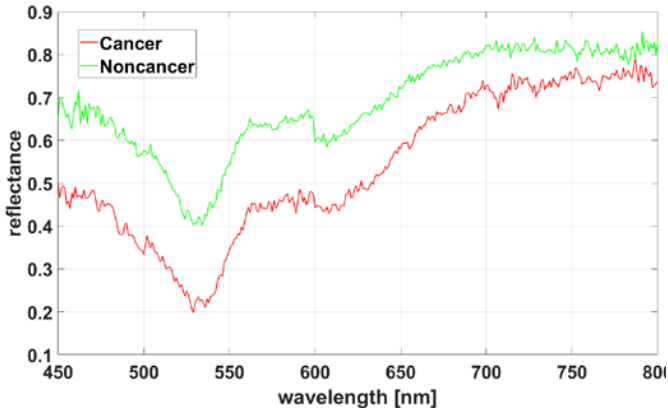

Fig. 2. Spectra of cancer-annotated pixel (red) and non-cancer-annotated pixel (green) in the 450-800 nm range, with a spectral resolution of 1 nm.

### C. Pixel-wise annotation

Pixel-wise annotation was performed by two pathologists, (A.P. and M.M.P.), and one medical expert (M.H.). As highlighted in section II.A, pixel-wise labeling of HSIs is challenging for pathologists due to the visual differences between HSIs and RGB images [9]. For example, in [10], HSIs and spatially aligned RGB images were acquired sequentially within the same field-of-view using a switch integrated into the microscope. The pixel-wise annotations made on RGB images were then transferred to the corresponding HSIs using custom-designed software. Since HSI system used in Fig. 1 cannot capture co-registered HSIs and RGB images simultaneously, we adopted the approach from [20]. For each HSI, we reconstructed a corresponding RGB image, a common practice in remote sensing for HSI visualization [44]. This process involves generating realistic color images by mapping the visible spectrum of the HSIs to the CIE XYZ color space [45], then converting it to the standardized RGB color space. The MATLAB code for method from [44] is publicly available at [46]. Subsequently, pixel-wise labeling by experts was conducted on the reconstructed RGB images, aided by a super-pixels-based software system, described in [1]. Fig. 3 shows an example of an RGB image reconstructed from an HSI, along with its ground truth map, generated by majority voting as explained below.

The annotation process is inherently affected by inter-observer and intra-observer variability, which can introduce subjectivity into the labeling of images [47]. To address this issue, multiple pathologists are involved in the annotation process to reduce these variabilities. After an initial round of annotations on RGB images, the same three experts were asked to repeat the annotation process three months later. To evaluate the reliability of the annotations and measure agreement among the annotators, Fleiss' kappa statistics were employed [48][49]. The agreement is: poor for kappa <0.00, slight for kappa between 0.00 and 0.20, fair for kappa between 0.21 and 0.40, moderate for kappa between 0.41 and 0.60, substantial for kappa between 0.61 and 0.80, and almost perfect for kappa between 0.81 and 1.00. Table I presents the kappa values indicating the agreement among three annotators (two pathologists and one medical expert) for the first and second labeling sessions. The agreement between the two pathologists was nearly perfect in both annotations, while their agreement with medical expert was significantly lower, particularly during the first annotation.

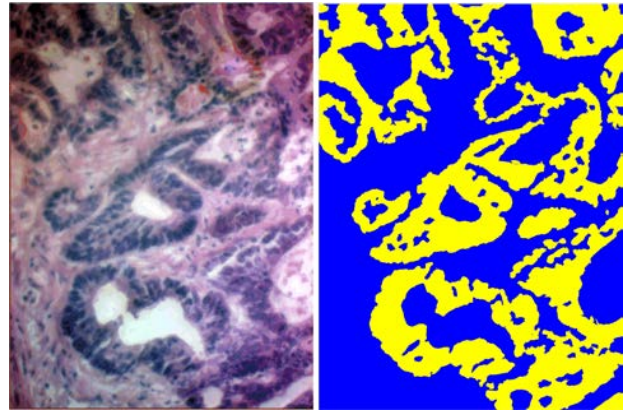

Fig. 3. Standardized RGB image (left) reconstructed from the corresponding HSI using the method from [44]. The associated ground truth (GT) map (right) shows pixel annotations, where yellow indicates cancerous tissue and blue represents non-cancerous tissue.

This confirms that the intra-annotator agreement for two pathologists was almost perfect or close to it. From Table I, it can be observed that the agreement between medical expert (M.H.) and the two pathologists improved in the second annotation session. For the formation of the final ground truth, we utilized both annotations from the pathologists and the second annotation from the medical expert through majority voting. Fig. 4 displays the annotations made by the two pathologists and the medical expert for the RGB image shown in Fig. 3. The source of disagreement between the pathologists and the medical expert stemmed from the categorization of a necrotic area within the tumor, which was addressed and resolved in the second annotation.

TABLE I
INTER-ANNOTATORS AGREEMENTS IN TERMS OF KAPPA STATISTICS FOR TWO ANNOTATION SESSIONS.

| Annot. | AP-MMP-MH | AP-MMP | AP-MH | MMP-MH |
|---|---|---|---|---|
| 1 | 0.6966 | 0.7698 | 0.6542 | 0.6636 |
| 2 | 0.7228 | 0.7687 | 0.6827 | 0.7163 |

### D. HSI classifiers based on spectral-spatial features

As is discussed in Section II.A, designing a deep architecture that effectively extracts discriminative spectral-spatial features from HSIs while reducing the need for a large amount of annotated pixels for training is challenging. As noted in [4], it is also crucial that the proposed architecture demonstrates a significant improvement in classification performance using HSIs compared to RGB images. These challenges are consistent with the experience in the remote sensing community, where deep-network-based HSI classification methods either offer only marginal improvements over hand-crafted features or require significantly more labeled data [50].

*1) The SSL classifier*

To reduce the reliance on a large amount of annotated pixels for training, we employed a SSL approach [67]. SSL algorithms aim to train a classifier and assign labels to unlabeled data, starting with a classifier trained on a small labeled dataset. Specifically, we used a self-training SSL algorithm [68], which assigns pseudo-labels to high-confidence unlabeled samples. These samples are then added to the training set, and the classifier is iteratively improved. In our implementation of this SSL classifier, we utilized the MATLAB function `fitsemiself`.

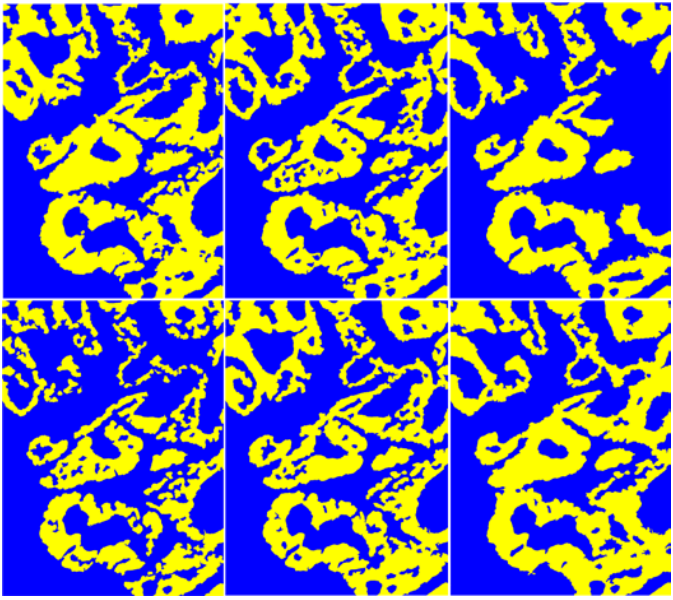

Fig. 4. Annotations from two pathologists and one medical expert are presented from left to right. Each set of annotations is shown from top to bottom: the first annotation followed by the second annotation. Yellow indicates pixels annotated as cancer, and blue indicates pixels annotated as non-cancer.

*2) The MPRI spectral-spatial features*

The MPRI architecture was recently introduced in hyperspectral remote sensing [35]. It learns discriminative spectral-spatial features for HSI classification by leveraging PRI [36]. For a random variable $\mathbf{X}$ with a known probability density function (PDF) $g$, PRI aims to learn a reduced statistical representation of $\mathbf{X}$, instantiated in a random variable $\mathbf{Y}$ with PDF $f$. This is framed as trade-off between the entropy $H(f)$ of $\mathbf{Y}$ and its descriptive power regarding $\mathbf{X}$, quantified by the divergence $D(f\|g)$ [36][35]:

$$\min_f H(f) + \beta D(f\|g). \quad (1)$$

Here, $\beta$ is a hyper-parameter that controls the amount of relevant information $\mathbf{Y}$ that can be extracted from $\mathbf{X}$. For a random variable $\mathbf{X}$, Reny's $\alpha$-entropy of $H_\alpha(\mathbf{X})$ can be used. When $\alpha=2$, the divergence in (1) can be expressed in terms of Cauchy-Schwartz (CS) divergence [51]:

$$D_{CS}(f\|g) = 2H_2(f;g) - H_2(f) - H_2(g). \quad (2)$$

The PRI optimization problem from (1) is then formulated as [35]:

$$f_{opt} \equiv \arg\min_f (1-\beta)H_2(f) + 2\beta H_2(f;g). \quad (3)$$

Let us assume $\mathbf{X} := \{\mathbf{x}_i\}_{i=1}^N$ and $\mathbf{Y} := \{\mathbf{y}_i\}_{i=1}^N$ are realizations of random variables drawn independently from $g$ and $f$, respectively. Using a Parzen's window-density estimator with a Gaussian kernel $G_\sigma(\cdot) = \exp(-\|\cdot\|^2/2\sigma^2)$ [52], equation (3) can be written as [53]:

$$\mathbf{Y}_* = \arg\min_\mathbf{Y} \begin{bmatrix} -(1-\beta)\log\left(\frac{1}{N^2}\sum_{i,j=1}^N G_\sigma(\mathbf{y}_i - \mathbf{y}_j)\right) \\ -2\beta\log\left(\frac{1}{N^2}\sum_{i,j=1}^N G_\sigma(\mathbf{y}_i - \mathbf{x}_j)\right) \end{bmatrix}. \quad (4)$$

Now, let us assume the HSI is given as a 3D data cube $\mathcal{T} \in \mathbb{R}^{r \times c \times d}$, $r$ and $c$ represent the number of rows and columns (spatial dimensions), and $d$ represents the number of spectral bands. Around each target vector $\mathbf{t}_* \in \mathbb{R}^d$, a local patch $\mathbf{T} \in \mathbb{R}^{N \times d}$ is extracted using a sliding window of size $n$, centered at $\mathbf{t}_*$. The total number of pixels in the patch is $N=n\times n$, and $\mathbf{t}_* = \mathbf{T}_{\lfloor n/2 \rfloor+1, \lfloor n/2 \rfloor+1}$, where $\lfloor \cdot \rfloor$ denotes the nearest integer function [35]. The PRI-based spectral-spatial characterization $\mathbf{Y}_* \in \mathbb{R}^{N \times d}$ of $\mathbf{T}$ is then computed using equation (4), making the estimation fully data driven. Section 3.1 in [35] describes the update equation for $\mathbf{Y}_*$. The PRI-based representation for the current patch is given by $\mathbf{t}_* = \mathbf{Y}_{\lfloor n/2 \rfloor+1, \lfloor n/2 \rfloor+1}$, and scanning the entire 3D cube $\mathcal{T}$ yields the PRI-based spectral-spatial representation $\mathcal{Y} \in \mathbb{R}^{r \times c \times d}$. To obtain PRI at multiple scales, procedure can be repeated for patches of varying sizes $n$, resulting in a multiscale PRI. The patch-size $n$ is a hyper-parameter. To reduce feature redundancy, regularized linear discriminant analysis is applied [54][35]. A multilayer structure is created by feeding PRI-based spectral-spatial features from one layer into the input the next. The final spectral-spatial representation is obtained by concatenating the representations from each layer. In the original work [35], this representation is used with a standard $k$-nearest neighbor (k-NN) classifier. However, in our contribution, we use an SSL classifier, as it achieves comparable performance with significantly fewer labeled samples. The MPRI architecture has four hyper-parameters that need tuning during cross-validation: the number of scales (determined by the patch size $n$), $\beta$, the number of layers, and the kernel variance $\sigma^2$ [35]. In our implementation, after cross-validation, we selected three layers, $n\in\{3, 7, 11\}$, $\beta\in\{2, 3\}$ and $\sigma^2=0.3$. The initial MATLAB-based implementation of MPRI is available at [55]. The MPRI architecture discriminates between tumor and stroma in HSI, whereas MPRI-based classification of the RGB images fails to do so. With 1% of labeled pixels per class, the SSL-MPRI classifier achieves a 4.87% increase in balanced accuracy and 5.31% increase in Dice coefficient.

### 3) The TensorSSA spectral-spatial features

Tensor Singular Spectrum Analysis (TensorSSA) [64], with MATLAB code in [65], is designed to extract global and low-rank 3D spectral-spatial features from HSI. The low-rank nature of extracted features is guaranteed through the tensor singular value decomposition (t-SVD) model [66]. The TensorSSA method first performs adaptive embedding of the HSI $\mathcal{T} \in \mathbb{R}^{r \times c \times d}$ onto a tensor $\mathcal{Z} \in \mathbb{R}^{l \times rc \times d}$, where $w \times w$ is the size of the patch centered around each pixel, with $w=2u+1$. $l \leq (w-2)^2$ is the number of pixels surrounding the central pixel that are evaluated for similarity to the central pixel, based on the normalized Euclidean distance. Both $u$ and $l$ are hyperparameters of the TensorSSA method. In our experiments reported in Section IV, after cross-validation we set $u=5$ and $l=8$ for processing RGB images, and $u=5$ and $l=60$ for processing HSIs. The trajectory tensor $\mathcal{Z}$ contains both spectral and spatial information corresponding to the entire HSI. The tensor $\mathcal{Z}$ is then replaced by its t-SVD-based low-tensor-tubal-rank approximation, i.e., $\mathcal{Z} \rightarrow \mathcal{Z}_{rtub} \in \mathbb{R}^{l \times rc \times d}$, where tubal rank satisfies $rtub << \min(l, rc)$. Experiments carried out in [64] showed that $rtub=1$ works well. The tensor $\mathcal{Z}_{rtub}$ is subsequently reprojected to $\mathcal{Y} \in \mathbb{R}^{r \times c \times d}$, which contains 3D spectral-spatial features associated with $\mathcal{T}$ [66]. While in [66] a linear support vector machine (SVM) algorithm was used to classify HSI based on these spectral-spatial features $\mathcal{Y}$, in this paper, we employ an SSL classifier, as we did for the MPRI spectral-spatial features, due to its effectiveness with fewer labeled samples.

Fig. 5 presents the classification results from SSL-MPRI classifier trained with 1% of labeled pixels per class, and SSL-TensorSSA classifier trained with 2% of labeled pixels per class, for the HSI and its corresponding RGB image shown in Fig. 3. It can be observed that SSL-MPRI classifier on the HSI correctly classifies stroma tissue as non-cancerous, a distinction not made by the other three cases.

### E. Deep learning HSI classifiers

As discussed in Section II.A, significant efforts have been made to design architectures that can effectively extract spectral-spatial features from HSIs, while reducing the need for large amount of annotated pixels for training. Several specialized architectures, such as S³-R [9], HyperNet [10], IANet [22], and transformers [27], have been proposed for this purpose. These architectures address the limited ability of U-shaped networks to model long-range contextual relationship across the spectral dimension of HSIs [10][11]. One such network termed U-within-U-Net (UwU-Net), was introduced in [56] as a convolutional framework that preserves spatial resolution and accommodates an arbitrary number of spectral channels. The code for UwU-Net is available at [57]. In Section IV, we apply UwU-Net to the segmentation of HSIs and RGB images of adenocarcinoma of the colon in the liver. Another network, nnUnet, was proposed in [58], with code in Supplementary Software link, to address the limitations of existing deep-learning-based methods for segmenting biomedical datasets, particularly the dependency of performance on dataset characteristics and hardware conditions. nnUnet automatically configures itself for a new segmentation task, and since it was trained on a large and diverse data pool, it is expected to perform well on datasets with limited training data. Therefore, we also applied nnUnet to the segmentation HSIs and RGB images of adenocarcinoma of the colon in the liver. Additionally, to corroborate some previous findings, we applied well-known deep networks for the classification of both HSIs and corresponding RGB images: DeepLabv3+ [59], Unet [60], Unet++ [61], and MA-Net [62]. MA-Net, in particular, incorporates a self-attention mechanism and is expected to effectively capture both spatial and spectral feature dependencies.

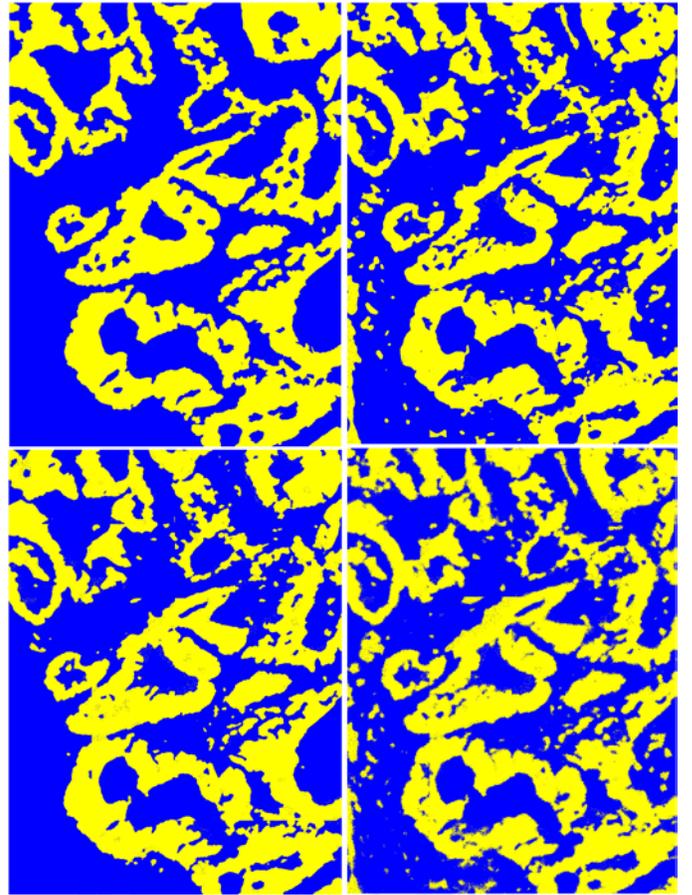

Fig. 5. Results of SSL-MPRI classifier (top) and SSL-TensorSSA classifier (bottom) on HSI (left) and corresponding RGB image, shown in Fig.3, (right). Ground truth image is also shown in Fig. 3. Yellow color denotes pixels classified as cancerous, blue color denotes pixels classified as non-cancerous. Number of annotated pixels per class is 1% for SSL-MPRI classifier and 2% for SSL-TensorSSA classifier. Balanced accuracy in respective order is: 1.000, 0.8604 (top), and 0.8878, 0.8699 (bottom). Value of Dice coefficient in respective order is: 1.0000 and 0.8386 (top), and 0.8699 and 0.8322 (bottom).

### F. Train and test protocols

*1) The SSL, k-NN and SVM classifiers combined with MPRI and TensorSSA 3D spectral-spatial features*

Due to the memory constraints, the SSL-MPRI and SSL-TensorSSA classifiers operated on input patches of size 230×258 pixels, meaning the HSIs and RGB images were divided into 24 patches. As a result, these classifiers work on a patch-by-patch basis, allowing them to handle cases where only a single image is available in a given scenario. This approach is often applied in deep learning models form remotely sensed HSIs [50].

*2) Deep learning classifiers*

For the deep learning classifiers, we split the datasets, consisting of HSIs and RGB images into 63% for training (17 images) and 37% for testing (10 images). To ensure fair performance validation, we made sure that the test set images came from different patients than the training set. In other words, the test set included five patients, none of whom were part of the training set, which contained images from nine patients. Regarding the deep networks used in the experiment (Unet, Unet++, DeepLabv3+, MA-Net, nnUnet, and UwU-Net) pixel values were normalized to a range between 0 and 1. We also applied a patching process using a 128×128 window with strides of 64 pixels. Additionally, 20% of the preprocessed train set was set aside as a validation set.

### G. Performance measures

In our experiments, reported in Section IV, we used six metrics to evaluate classification performance of different architectures on HSIs and their corresponding RGB images: sensitivity (SE), also known as recall or true positive rate; specificity (SP), also called selectivity or true negative rate; balanced accuracy (BACC); precision (PREC), also known as positive predicted value; $F_1$ score (or Dice coefficient); and intersection-over-union (IoU), also referred to as the Jaccard index. Each metric ranges from 0 to 1, where 0 indicates the worst performance and 1 indicates the best. All metrics are defined in terms of true positives (TP: the number of correctly identified cancerous pixels), true negatives (TN: the number of correctly identified non-cancerous pixels), false positives (FP: the number of incorrectly identified non-cancerous pixels), and false negatives (FN: the number of incorrectly identified cancerous pixels). The metrics are calculated using the following formulas:

$$SE = \frac{TP}{TP+FN}, \quad SP = \frac{TN}{TN+FP}, \quad BACC = \frac{SE+SP}{2}$$

$$F_1 = \frac{2TP}{2TP+FP+FN}, \quad IoU = \frac{TP}{TP+FN+FP}$$

$$PREC = \frac{TP}{TP+FP}. \quad (5)$$

## IV. EXPERIMENTS AND RESULTS

We conducted experiments to determine whether the selected methods: (*i*) provide significantly better classification of cancerous and non-cancerous pixels from HSIs compared to corresponding RGB images (in terms of statistical significance), and (*ii*) reduce the need for a large number of labeled pixels necessary for training.

### A. Software environment

We implemented the SSL-MPRI, SSL-TensorSSA, k-NN-MPRI and SVM-TensorSSA classifiers in MATLAB on a computer running a 64-bit Windows 10 operating system. The computer was equipped with 256 GB of RAM and an Intel Xeon CPU E5-2650 v4 2 processor, operating at a clock speed of 2.2 GHz. All deep learning-based classifiers were implemented using the PyThorch software environment [63].

### B. Results

We first conducted a study to determine whether the SSL-MPRI and SSL-TensorSSA classifiers should be preferred over the k-NN-MPRI, SVM-TensorSSA and standard SSL-spectral classifiers for both HSIs and corresponding RGB images. For each patch, 1% and 2% of pixels per class were labeled to train the SSL-based classifiers. For comparison, k-NN-MPRI and linear SVM-TensorSSA classifiers were trained using 20% and 5% of labeled pixels per class, respectively. The results of our study are presented in Table II, focusing on micro performance. For both HSI-s and corresponding RGB images, the SSL-MPRI classifier, using only 2% labeled pixels per class, achieved the highest BACC, $F_1$ score, and IoU. This performance was superior to the k-NN-MPRI classifier and the SVM-TensorSSA classifier. It also outperformed the SSL-spectral classifier applied to both HSIs and RGB images. These results justify the use of the SSL-MPRI classifier over the k-NN-MPRI classifier, as well as the use of HSIs over RGB images, given the significantly better performance achieved on HSIs. We show in Table III, performance of SSL-MPRI and SSL-TensorSSA classifiers using only 1% of labeled pixels per class for training. For SSL-MPRI and SSL-TensorSSA classifiers, we report in Table IV performance metrics as mean±standard deviation, calculated on an image-by-image basis. This approach allows for the evaluation of statistical significance between performance on HSIs and their corresponding RGB images. Additionally, we assessed statistical significance between the two classifiers on HSIs using the Wilcoxon rank-sum test. The null hypothesis of the test assumes that the data come from continuous distributions with equal medians at a 5% significance level. A p-value less than 0.05 indicates rejection of the null hypothesis. As shown, both classifiers achieve statistically significant improvements on HSIs compared to RGB images across all performance metrics. On HSIs, the SSL-MPRI classifier also achieves a statistically significant performance improvement over the SSL-TensorSSA classifier. For deep networks, we provided in Table V the mean and standard deviation obtained over 10 runs. The best performance overall is achieved by the nn-Unet, but there is no difference between HSIs and RGB images. In comparison with the SSL-MPRI classifier the performance is, depending on the metrics, 2% to 8% worse.

TABLE II
MICRO PERFORMANCE METRICS FOR THE K-NN-MPRI, SVM-TENSORSSA, SSL-SPECTRAL, SSL-MPRI AND SSL-TENSORSSA ALGORITHMS VS. PERCENTAGE OF LABELED PIXELS-PER-CLASS. EVALUATION ON 27 HSIS AND RGB IMAGES. PATCH SIZE 230×258 PIXELS.

| Algorithm | Perc./Image | | SE | SP | BACC | $F_1$ | IoU | PREC. | CPU [min/patch] |
|---|---|---|---|---|---|---|---|---|---|
| k-NN-MPRI | 20 | HSI | 0.9120 | 0.9339 | 0.9265 | 0.9188 | 0.8498 | 0.9186 | 56.72 |
| | | RGB | 0.8621 | 0.8873 | 0.8747 | 0.8617 | 0.7570 | 0.8612 | 15.95 |
| SVM-TensorSSA | 5 | HSI | 0.8918 | 0.8958 | 0.8938 | 0.8828 | 0.7902 | 0.8740 | 3.38 |
| | | RGB | 0.8860 | 0.8886 | 0.8873 | 0.8757 | 0.7790 | 0.8658 | 0.10 |
| SSL-spectral | 2 | HSI | 0.9109 | 0.9023 | 0.9066 | 0.8965 | 0.8124 | 0.8826 | 17.27 |
| | | RGB | 0.8700 | 0.8540 | 0.8620 | 0.8483 | 0.7366 | 0.8227 | 2.07 |
| SSL-MPRI | 2 | HSI | 0.9435 | 0.9299 | 0.9363 | 0.9289 | 0.8672 | 0.9147 | 16.07 |
| | | RGB | 0.8871 | 0.8780 | 0.8826 | 0.8704 | 0.7705 | 0.8543 | 8.18 |
| SSL-TensorSSA | 2 | HSI | 0.9159 | 0.9067 | 0.9113 | 0.9016 | 0.8209 | 0.8878 | 15.75 |
| | | RGB | 0.8700 | 0.8584 | 0.8680 | 0.8548 | 0.7465 | 0.8332 | 1.75 |

TABLE III
MICRO PERFORMANCE METRICS FOR THE SSL-MPRI AND SSL-TENSORSSA CLASSIFIERS FOR 1% OF LABELED PIXELS-PER-CLASS. EVALUATION ON 27 HSIS AND RGB IMAGES. PATCH SIZE 230×258 PIXELS.

| Algorithm | Image | SE | SP | BACC | $F_1$ | IoU | PREC. | CPU [min/patch] |
|---|---|---|---|---|---|---|---|---|
| SSL-MPRI | HSI | 0.9399 | 0.9227 | 0.9313 | 0.9235 | 0.8578 | 0.9076 | 29.60 |
| | RGB | 0.8871 | 0.8780 | 0.8826 | 0.8704 | 0.7705 | 0.8543 | 11.82 |
| SSL-TensorSSA | HSI | 0.9111 | 0.9000 | 0.9055 | 0.8955 | 0.8108 | 0.8804 | 10.25 |
| | RGB | 0.8746 | 0.8540 | 0.8643 | 0.8511 | 0.7790 | 0.8289 | 1.48 |

TABLE IV
MACRO PERFORMANCE METRICS FOR THE SSL-MPRI AND SSL-TENSORSSA CLASSIFIERS FOR 1% OF LABELED PIXELS-PER-CLASS. EVALUATION ON 27 HYPERSPECTRAL IMAGES (HSIS) AND CORRESPONDING RGB IMAGES. WILCOX RANK-SUM TEST OF STATISTICAL SIGNIFICANCE WITHIN 95% CONFIDENCE INTERVAL (P-VALUE). LAST ROW REPRESENTS STATISTICAL SIGNIFICANCE BETWEEN SSL-MPRI AND SSL-TENSORSSA CLASSIFIERS ON HSIS.

| Algorithm | Image | SE | SP | BACC | $F_1$ | IoU | PREC. |
|---|---|---|---|---|---|---|---|
| SSL-MPRI | HSI | 0.9402±0.0131 | 0.9177±0.0256 | 0.9290±0.0153 | 0.9210±0.0141 | 0.8539±0.0244 | 0.9029±0.0232 |
| | RGB | 0.8877±0.0234 | 0.8736±0.0254 | 0.8806±0.0190 | 0.8616±0.0374 | 0.7587±0.0560 | 0.8405±0.0681 |
| p-value | | $5.28\times10^{-10}$ | $5.25\times10^{-7}$ | $8.78\times10^{-10}$ | $1.94\times10^{-9}$ | $1.94\times10^{-9}$ | $2.43\times10^{-3}$ |
| SSL-TensorSSA | HSI | 0.9128±0.0204 | 0.8948±0.0281 | 0.9038±0.0196 | 0.8918±0.0206 | 0.8054±0.0335 | 0.8729±0.0370 |
| | RGB | 0.8772±0.0294 | 0.8507±0.0305 | 0.8639±0.0243 | 0.8435±0.0429 | 0.7316±0.0617 | 0.8165±0.0744 |
| p-value | | $3.85\times10^{-6}$ | $4.45\times10^{-5}$ | $1.07\times10^{-7}$ | $4.19\times10^{-6}$ | $4.19\times10^{-6}$ | $2.00\times10^{-3}$ |
| p-value | | $8.20\times10^{-7}$ | $3.70\times10^{-3}$ | $1.03\times10^{-5}$ | $2.14\times10^{-6}$ | $2.14\times10^{-6}$ | $4.60\times10^{-3}$ |

TABLE V
MICRO PERFORMANCE METRICS FOR THE DEEP-NEURAL NETWORKS (DNN) AVERAGED OVER 10 RUNS. EVALUATION ON 10 HYPERSPECTRAL AND RGB IMAGES FROM THE TEST SET.

| DNN | Image | SE | SP | BACC | $F_1$ | IoU | PREC. |
|---|---|---|---|---|---|---|---|
| DLabv3+ | HSI | 0.8955±0.0245 | 0.8287±0.0477 | 0.8621±0.0192 | 0.8517±0.0188 | 0.7421±0.0283 | 0.8136±0.0397 |
| | RGB | 0.8805±0.0207 | 0.8527±0.0224 | 0.8667±0.0027 | 0.8553±0.0027 | 0.7422±0.0041 | 0.8321±0.0189 |
| Unet | HSI | 0.8876±0.0733 | 0.7691±0.1140 | 0.8283±0.0368 | 0.8197±0.0333 | 0.6958±0.0484 | 0.7726±0.0863 |
| | RGB | 0.8671±0.0140 | 0.8655±0.0181 | 0.8663±0.0055 | 0.8542±0.0056 | 0.7455±0.0085 | 0.8420±0.0160 |
| Unet++ | HSI | 0.9006±0.0907 | 0.7839±0.0578 | 0.8423±0.0298 | 0.8309±0.0370 | 0.7123±0.0522 | 0.7777±0.0368 |
| | RGB | 0.8531±0.0269 | 0.8792±0.0164 | 0.8661±0.0079 | 0.8531±0.0098 | 0.7439±0.0146 | 0.8539±0.0137 |
| MAnet | HSI | 0.9193±0.0224 | 0.8004±0.0341 | 0.8559±0.0120 | 0.8505±0.0112 | 0.7401±0.0168 | 0.7923±0.0254 |
| | RGB | 0.8510±0.0023 | 0.8761±0.0126 | 0.8635±0.0054 | 0.8504±0.0059 | 0.7398±0.0089 | 0.8501±0.0120 |
| UwU-Net | HSI | 0.9520±0.0367 | 0.7334±0.0650 | 0.8427±0.0210 | 0.8369±0.0177 | 0.7200±0.0259 | 0.7490±0.0412 |
| | RGB | 0.7913±0.0557 | 0.8639±0.0508 | 0.8276±0.0121 | 0.8085±0.0168 | 0.6789±0.0221 | 0.8312±0.0386 |
| nn-Unet | HSI | 0.9212±0.0209 | 0.8595±0.0491 | 0.8904±0.0293 | 0.8654±0.0832 | 0.7627±0.1124 | 0.8272±0.1261 |
| | RGB | 0.8852±0.0463 | 0.9010±0.0291 | 0.8913±0.0297 | 0.8716±0.0463 | 0.7724±0.0613 | 0.8649±0.0844 |

## V. Discussion

Efficacy of intraoperative tumor resection directly affects life expectancy, i.e., the resection margin has been known as a powerful predictor of the 5-year survival rate. Traditionally, computational pathology has relied on RGB images of frozen tissue sections, which limits information to the visual range and excludes the data across the continuous spectral range. To address this limitation, HSI technology is increasingly being used in various medical imaging applications, including computational pathology. However, application of HSI in computational pathology faces three challenges: (1) a shortage of pixel-wise annotated HSI data necessary for training machine learning and DL models; (2) in many cases, the advantages of HSI over traditional RGB imaging have not been conclusively demonstrated, particularly for specimens collected intraoperatively; (3) experimental variability in slide preparation process, also known as batch effects [69] [70], transfers to spectrum variability and that, in combination with the lack of annotated data, makes training of deep networks challenging. To address first two challenges we created a database consisted of 27 HSIs of H&E stained frozen sections, collected from 14 patients with colon adenocarcinoma metastasized to the liver. The ground-truth data based on pixel-wise annotations performed by three pathologists are also available. To overcome third challenge, we combined label-propagation-based SSL with spectral-spatial features extracted by the multiscale principle of relevant information (MPRI) method and tensor singular spectrum analysis method. While proposed approach achieved highly competitive performance in intraoperative tumor demarcation problem it still demands involvement of the human expert to label a small amount pixels as cancerous and non-cancerous (1% per class). Our future efforts will be directed towards fully automation of the tumor demarcation problem through usage of some state-of-the-art subspace clustering methods, [71], to generate initial small amount of high-quality pseudo labels for SSL-MPRI classifier.

## VI. Conclusion

In this work, we addressed intraoperative tumor resection problem through pixel-wise classification of HSIs of H&E stained specimens of metastatic colon cancer in the liver. By using only 1% of labeled pixels per class (cancer vs. non-cancer) generated locally (on the patch level), the SSL-MPRI classifier achieved a micro balanced accuracy (BACC) of 0.9313 and a micro F1-score of 0.9235 on the HSI dataset. In comparison, the performance on corresponding RGB images was lower, with a micro BACC of 0.8809 and a micro F1-score of 0.8688. The improvements relative to RGB images are statistically significant in six classification performance metrics. The SSL-MPRI approach also outperformed six DL architectures trained with 63% of labeled pixels. Moreover, DL architectures achieved on HSIs classification performance no better than the one on corresponding RGB images. We conclude that SSL-MPRI classifier in combination with HSI represents state-of-the-art solution for intraoperative computation of tumor resection margin.